\documentclass[a4paper,pre,twocolumn,longbibliography]{revtex4-1}
\usepackage[dvips]{graphicx}
\usepackage{amsmath}
\usepackage{psfrag,color}
\usepackage{amssymb}
\usepackage{natbib}
\usepackage{ulem}
\usepackage{mathtools}
\usepackage{siunitx}
\usepackage{pifont}
\usepackage{bm}

\newcommand{\myeqref}[2]{Eq.~(\hyperref[#1]{\ref*{#1}#2})}
\newcommand{\myref}[2]{\hyperref[#1]{\ref*{#1}(#2)}}
\newcommand{\myrefletter}[2]{\hyperref[#1]{(#2)}}
\newcommand{\myrefnb}[2]{\hyperref[#1]{\ref*{#1}#2}}

\newcommand{\srr}{\sigma_{rr}}
\newcommand{\stt}{\sigma_{\theta\theta}}
\newcommand{\sttmin}{\sigma_{\theta\theta}^{\mathrm{min}}}
\newcommand{\tW}{\tilde{W}}
\newcommand{\tWc}{\tilde{W}_c}
\newcommand{\tWcomp}{\tilde{W}_{\mathrm{comp}}}
\newcommand{\tWdewet}{\tilde{W}_{\mathrm{dewet}}}

\newcommand{\lbc}{\ell_{bc}}
\newcommand{\DeltaTot}{\Delta_{\mathrm{tot}}}
\newcommand{\Ubend}{U_{\mathrm{bend}}}

\newcommand{\uribbon}{u_{\mathrm{rib}}}
\newcommand{\Ustrain}{U_{\mathrm{strain}}}
\newcommand{\nruck}{n_{\mathrm{ruck}}}
\newcommand{\nfold}{n_{\mathrm{fold}}}
\newcommand{\fold}{\mathrm{fold}}
\newcommand{\ruck}{\mathrm{ruck}}

\newcommand{\conf}{\mathcal{C}}

\newcommand{\dd}{\text{d}}

\begin{document}
%----- Title:
\title{Delamination from an adhesive sphere: Curvature--induced dewetting versus buckling}
%----- Authors
% Use letters for affiliations, numbers to show equal authorship (if applicable) and to indicate the corresponding author
\author{Finn Box$^{a,b}$, Lucie Domino$^{a,c}$ , Tiago Outerelo Corvo$^a$, Mokhtar Adda-Bedia$^d$, Vincent D\'{e}mery$^{d,e}$, Dominic Vella$^{a}$ and Benny Davidovitch$^f$}
%\author[a,2]{}
%\author[f]{}

\affiliation{$^a$Mathematical Institute, University of Oxford, Woodstock Rd, Oxford, OX2 6GG, UK\\
$^b$Department of Physics and Astronomy, University of Manchester, Manchester M13 9PL, UK\\
$^c$Institute of Physics, Universiteit van Amsterdam, Science Park 904, 1098 XH, Amsterdam, The Netherlands\\
$^d$Univ Lyon, ENS de Lyon, CNRS, Laboratoire de Physique, 69342 Lyon, France\\
$^e$Gulliver, CNRS, ESPCI Paris PSL, 10 rue Vauquelin, 75005 Paris, France\\
$^f$Department of Physics, University of Massachusetts, Amherst, MA 01003, USA}

\date{\today}

\begin{abstract}
Everyday experience confirms the tendency of adhesive films to detach from spheroidal regions of rigid substrates --- what is a petty frustration when placing a sticky bandage onto an elbow or knee is a more serious matter in the coating and painting industries. Irrespective of  their resistance to bending, a key driver of
 such phenomena is Gauss' \textit{Theorema Egregium}, which implies that  naturally flat sheets cannot conform to doubly-curved surfaces without developing a strain whose magnitude grows sharply with the curved area. Previous attempts to characterize the onset of curvature-induced delamination, and the complex patterns it gives rise to, assumed a dewetting-like mechanism in which     the propensity of two materials to form contact through interfacial energy is modified by an elastic energy penalty. We show that this approach may characterize moderately bendable adhesive sheets, but fails qualitatively to describe the curvature-induced delamination of ultrathin films, whose mechanics is governed by 
    their propensity to buckle under minute levels of compression. Combining mechanical and geometrical considerations, we introduce a minimal model for curvature-induced delamination that accounts for two elementary buckling motifs, shallow ``rucks'' and localized ``folds''. We predict nontrivial scaling rules for the onset of curvature-induced delamination and various features of the emerging patterns, which compare well with experimental observations. Beyond gaining control on the use of ultrathin adhesives in cutting edge technologies such as stretchable electronics, our analysis is a significant step towards quantifying the multiscale morphological complexity that emerges upon imposing geometrical and mechanical constraints on highly bendable solid objects.       
\end{abstract}

\maketitle

\section{Introduction}

The simplest way to form a composite material, simply sticking a layer of one material to another,  is encountered in everyday life from sticky notes to a parent placing a band-aid on a child's knee. Normally one does not consider whether such an operation is at all possible. However, when at least one of the two objects to be joined is  curved, adhesion is no longer guaranteed. For example, when a relatively thick, flat plate is adhered to a cylindrical substrate the energetic penalty associated with detaching from the substrate is small enough to be overcome by the elastic (bending) energy of the plate that is released by detachment  \cite{Neukirch2007} and leads to the failure of the coating, or delamination. For sufficiently thin plates, this bending energy is insignificant and adhesion proceeds as expected.

\begin{figure}
\begin{center}
\includegraphics[width=.95\columnwidth]{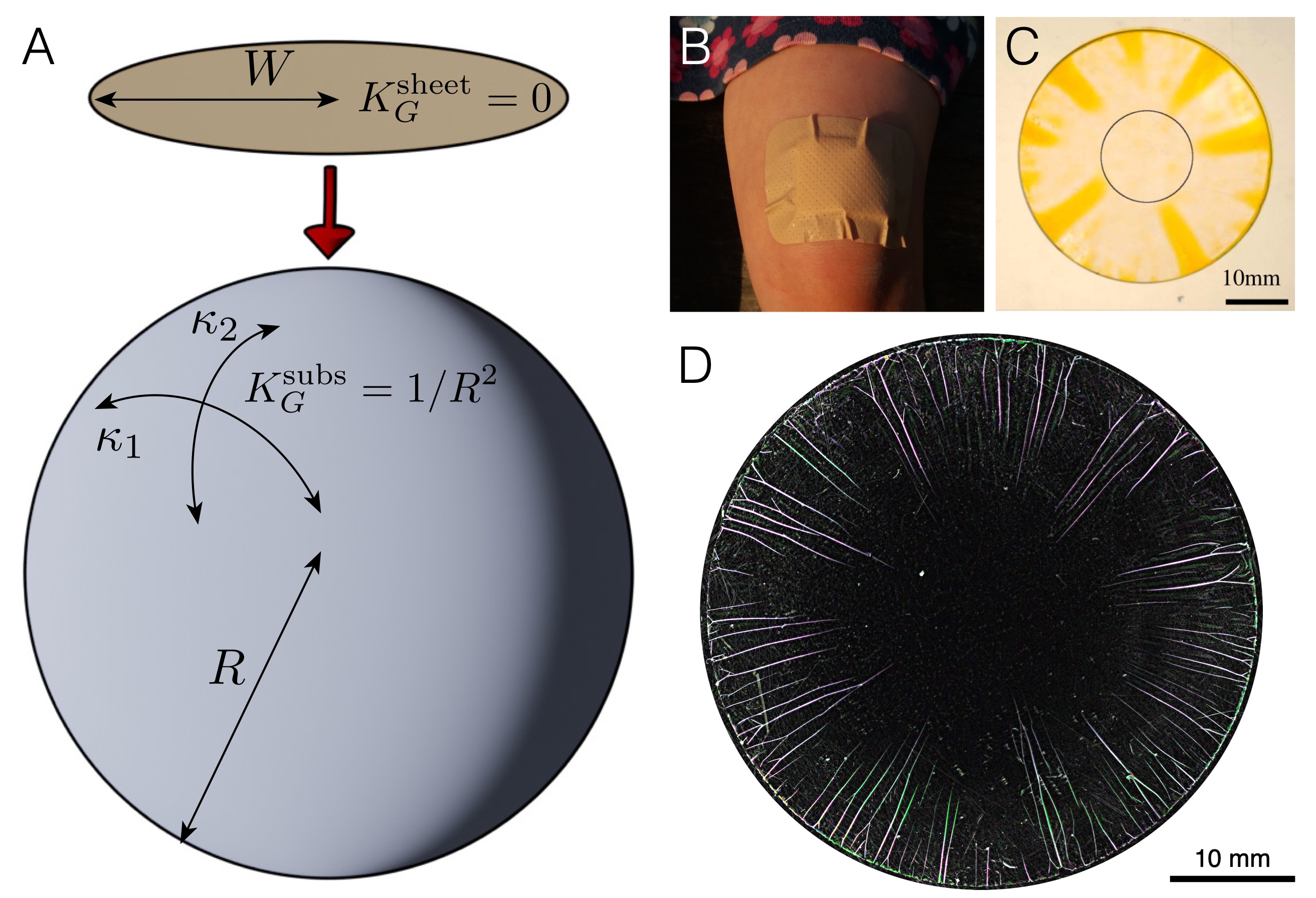}
\end{center}
\caption{Illustrations of the geometric incompatibility of an intrinsically flat sheet adhering to a doubly-curved object. A: A circular, naturally flat, sheet of radius $W$ has Gaussian curvature $K_G^{\mathrm{sheet}}=0$ while a spherical substrate of radius $R$ has Gaussian curvature $K_G^{\mathrm{subs}}=\kappa_1\kappa_2=1/R^2$. B: Delamination blisters spontaneously form when a (flat) band-aid is adhered to a child's (curved) knee. Panels C and D show more controlled realizations of this experiment in sheets of large and small thicknesses, respectively. (Image in C taken from \cite{Hure2011}.) }
\label{fig:Setup}
\end{figure}

A qualitatively different picture emerges when the substrate is doubly-curved (i.e.~has two principal curvatures $\kappa_{1,2}\neq0$, fig.~\ref{fig:Setup}A). While very thin sheets are able to bend easily, stretching is much more difficult and so Gauss' \emph{Theorema Egregium} \cite{Wilson2008} limits them to maintaining their initial Gaussian curvature, $K_G^{\mathrm{sheet}}$. If the Gaussian curvature of the substrate $K_G^{\mathrm{subs}}=\kappa_1\kappa_2\neq K_G^{\mathrm{sheet}}$ adhesion between the two frustrates the deformable object (the sheet). Ultimately this frustration can be (partially) relieved by delamination,  which allows the sheet to `expel' excess material and so conform to the substrate \cite{Paulsen2019,Vella2019}. The central result of this study is that this process may occur in two sharply distinct modes:  a dewetting-like instability (when the sheet is moderately bendable) or a buckling-like instability (for highly bendable sheets) in which most of the sheet remains adhered to the substrate and delamination occurs only at localized ``rucks'' or ``folds''.  
These distinctive modes of delamination make a significant difference to the maximal size of sheet that can be smoothly adhered before delamination occurs.

The geometrical frustration resulting from a difference in Gaussian curvature occurs when a flat sheet is deposited on a sphere, as is shown schematically in fig.~\ref{fig:Setup}A and in a practical scenario in fig.~\ref{fig:Setup}B. For the example shown in fig.~\ref{fig:Setup}B, this incompatibility leads to failure in the form of delamination blisters forming around the periphery. While this is a minor annoyance in the example of a band-aid applied to a curved knee, in technological and scientific applications conformability is key and so delamination is problematic \cite{Xu2019,Yuk2020,Tringides2021,Yan2022}. As a result, a variety of techniques have been developed to overcome geometric incompatibility ranging from modifying the substrate geometry (as in the Surface Force Apparatus \cite{Tabor1969,Israelachvili2015}) to buffering the excess area required to change the sheet's shape \cite{Vella2019} by either removing  material \cite{Cho2014,Liu2021,Yan2022} or introducing sacrificial buckling elements \cite{Wang2010,Jung2011}.

%\subsection{The standard picture of delamination}
Despite the broad significance of geometric incompatibility for adhesion, there seems to be little detailed understanding of \emph{when} and \emph{how} this incompatibility is expected to lead to failure via delamination. The standard picture of delamination induced by geometric incompatibility is due to Majidi \& Fearing \cite{Majidi2008}, who observed
that if a flat sheet (of radius $W$) is forced to adhere to a sphere of radius $R$, a strain of order $\varepsilon\sim K_G^{\mathrm{subs}}W^2\sim W^2/R^2$ is induced. Perfect adhesion therefore induces an elastic energy density $Y(W/R)^4$, where $Y=Et$ (with $E$ the Young modulus and $t$ the sheet thickness) is the stretching modulus of the sheet. Now, this elastic energy penalty can be avoided if the sheet retains a planar shape by completely detaching from the substrate, at the expense of paying an energetic penalty $\Gamma$ per unit area of lost contact. 

Assuming the sheet is either fully adhered to the substrate or fully delaminated from it, one may define a 
renormalized, curvature-dependent adhesion energy density: 
\begin{equation}
 \Gamma^* =    \Gamma - c \cdot Y(W/R)^4  \ ,  \label{eq:Gamma-ren}
\end{equation} 
where $c$ is a numerical constant that depends on the substrate geometry ($c= 1/384$
for a spherical substrate \cite{Grason2013,Hohlfeld2015}). 
Equation~(\ref{eq:Gamma-ren}) underlies an elementary description of delamination %can be understood 
as a generalization of the standard dewetting transition (which occurs as $\Gamma \to 0^+$), 
to ``curvature-induced dewetting'' (which occurs as $\Gamma^* \to 0^+$). In this generalization, the sheet's elastic energy is simply viewed as additional to the interfacial energies between the sheet, the substrate, and the surrounding phase. In terms of the dimensionless parameters:
\begin{equation}
 \tW = W/R \ \ ; \ \ \beta=\Gamma/Y    
\end{equation}
the curvature-induced dewetting scenario predicts that delamination occurs when $\tW$ exceeds a critical value: 
\begin{equation}
\tWdewet\sim \beta^{1/4}  \ . 
\label{eqn:WcritClassic}
\end{equation}

Although the curvature-induced dewetting mechanism does capture the basic competition between adhesion and geometrical constraints, considering $\Gamma^*$, defined in \eqref{eq:Gamma-ren}, as the single parameter that determines the onset of  
delamination is problematic. Indeed, treating adhesion and elastic energies as equivalent competitors obscures the fact that the former is uniformly distributed while the latter is distributed in a nontrivial and inhomogeneous manner: not only does the magnitude of strain vary significantly with radial distance from the center, but the spatial structure of its components is quite nontrivial. In particular, while the inner part of the sheet is stretched both radially and azimuthally, the periphery is stretched predominantly radially, and becomes azimuthally compressed when the sheet exceeds a critical size \cite{Grason2013,Hohlfeld2015}:      
\begin{equation}
    \tWcomp \sim \beta^{1/2} \ . 
    \label{eq:crit-2}
\end{equation}
Notably, for $\beta \ll 1$, $\tWcomp$ may be significantly smaller than the delamination size $\tWdewet$ predicted by the curvature-induced dewetting mechanism, Eq.~(\ref{eqn:WcritClassic}). Since the thinner a sheet is the less compression it can accommodate before buckling, one may expect that the onset of curvature-induced delamination in sufficiently thin sheets is not correctly described by the single parameter $\Gamma^*$, but requires an explicit consideration of the strong nonuniformity and anisotropy, and the consequent possibility of anisotropic instabilities such as radial wrinkling  \cite{Davidovitch2011}. The possible relevance of such a strongly anisotropic response to delamination is clear in fig.~\ref{fig:Setup}C,D, which shows that delamination occurs via the formation of radial blisters. In this paper, we show that an anisotropic instability of this type is crucial in understanding the threshold for delamination and that \eqref{eqn:WcritClassic} is, at best, valid only for sheets with moderate-to-large bendability.

\begin{figure}
\begin{center}
\includegraphics[width=.95\linewidth]{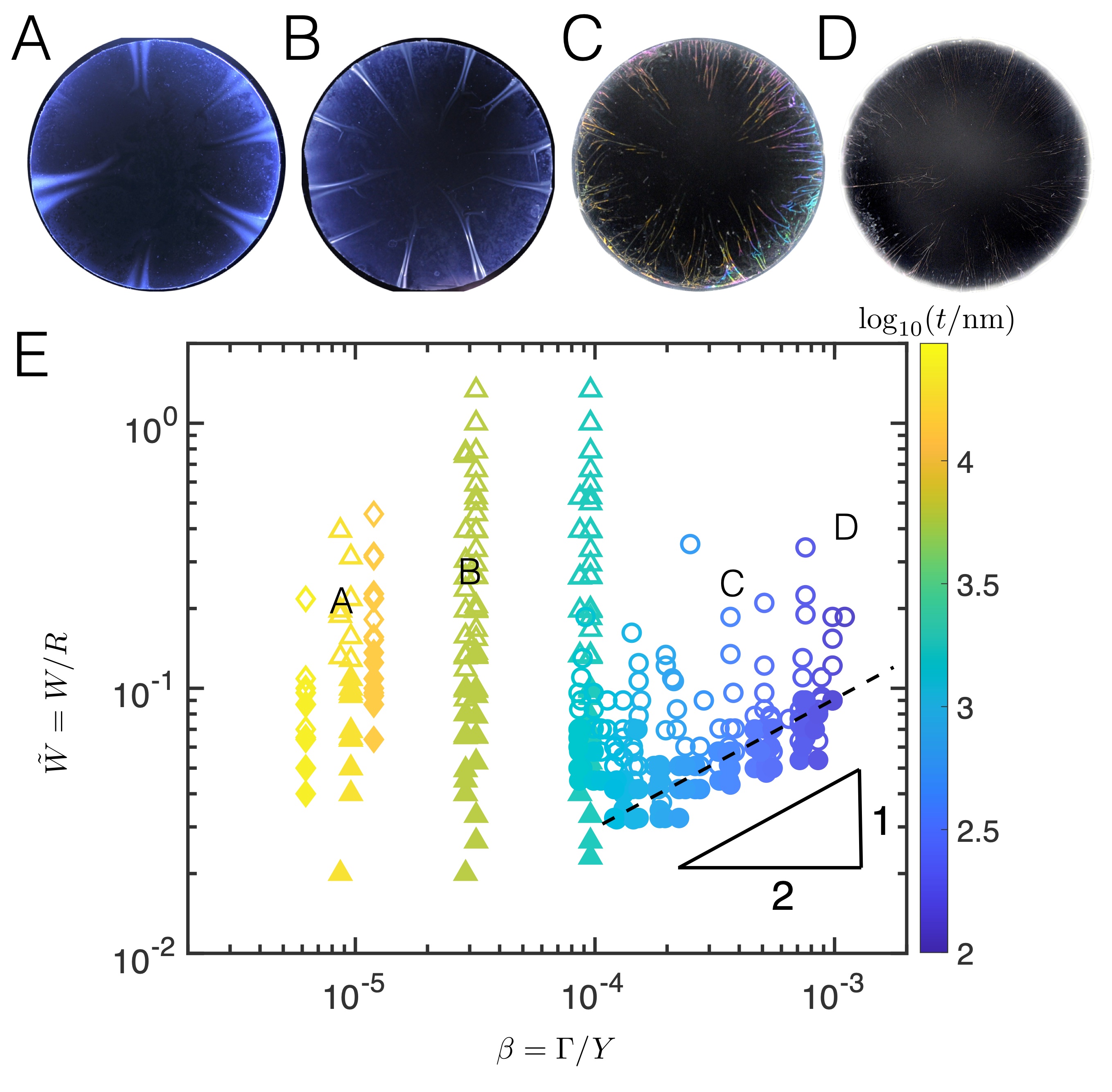}
\end{center}
\caption{A-D: Images showing typical delamination patterns for a range of sphere sizes and sheet thicknesses. E: Initial presentation of experimental results showing the regions of $(\beta,W/R=\tW)$ parameter space for which smooth adhesion (filled symbols) or localized delamination (open symbols) were observed. Contrary to previous suggestions, the delamination transition does \emph{not}  occur when  $\tWc\sim\beta^{1/4}$. (Determining the correct transition behavior is the focus of this paper.) Points are shown for a variety of sheet thicknesses, indicated by point color (see colorbar to the right), and  material, indicated by shape: Polyimide (diamond), Polycarbonate (triangles) and Polystyrene (circles). }
\label{fig:InitialRegimeDiagram}
\end{figure}

As a first indication that the energetic picture is not the full story, we present experimental results that interrogate the transition from smooth adhesion to delamination for circular sheets of radius $W$ and thickness $t$ and a sphere of radius $R$. In these experiments (see Supplementary Information), sheets (with thickness $t\in[100\mathrm{~nm},25\mathrm{~\mu m}]$) are deposited from floating on a water bath onto the sphere. After the system dries, the sheet is observed to either be smoothly adhered (represented by a filled point in fig.~\ref{fig:InitialRegimeDiagram}) or be partially delaminated (an open point in fig.~\ref{fig:InitialRegimeDiagram}). As expected, fig.~\ref{fig:InitialRegimeDiagram} shows that above a critical radius ratio $\tW>\tWc$ the sheet delaminates (modulo some imperfection-induced noise close to the transition).
Nevertheless, the  experimental data shown in fig.~\ref{fig:InitialRegimeDiagram} does not show the scaling
$\tWc\sim\beta^{1/4}$ predicted by the curvature-induced dewetting mechanism of \eqref{eqn:WcritClassic}: the only plausibly power-law behaviour that we observe appears to be $\tWc\sim\beta^{1/2}${, reminiscent of \eqref{eq:crit-2},} and this scaling is observed only in the very thinnest sheets.
%\rout{, in accord with \eqref{eq:crit-2}}. \dv{Not in accord, since the significance of compression is not yet clear.} 
This suggests that the simple theoretical picture of an energetic balance between stretching and adhesion, as exemplified by \eqref{eqn:WcritClassic}, is not relevant for this simple experimental system.  In this paper, we will show that  alternatives to the curvature-induced dewetting picture laid out by Majidi \& Fearing \cite{Majidi2008} exist; these do take into consideration the onset of compressive azimuthal stress, \eqref{eq:crit-2} and may actually be energetically favourable to the upper bound represented by \eqref{eqn:WcritClassic}.

\section{The onset of blistering}

\subsection{The importance of compression}

We begin our re-examination of the theory by taking a step back to consider the stress state within the elastic film as the sheet radius $W$ changes under the assumption that the sheet remains perfectly attached to the sphere, and hence axisymmetric. In this case, the vertical deformation of the sheet is 
\begin{equation}
\zeta=-r^2/(2R),
\label{eqn:SphereDefm}
\end{equation} (from the parabolic approximation to the sphere's surface, valid when $W/R\ll1$). Assuming that a tension $\Gamma$ acts at the edge of the sheet (originating from the adhesion), the stress profile within the sheet can then be readily calculated from this displacement, and has been presented in related work \cite{Grason2013,Hohlfeld2015}. This calculation gives that the radial and hoop stresses within the sheet induced by the deformation of \eqref{eqn:SphereDefm} are
\begin{align}
\srr&=\Gamma+\frac{Y}{16R^2}\left(W^2-r^2\right),
\label{eqn:LaminatedStressSrr}\\
\stt&=\Gamma+\frac{Y}{16R^2}\left(W^2-3r^2\right).
\label{eqn:LaminatedStressSqq}
\end{align}

Since $r\leq W$, we see that $\srr>0$ throughout the sheet, but also that the minimum value of the hoop stress, $\sttmin$ is
\begin{equation}
\sttmin=\stt(W) =\Gamma- \frac{YW^2}{8R^2}=Y \left(\beta-\frac{\tW^2}{8} \right),
\label{eqn:sttLaminated}
\end{equation} while the radial displacement of the sheet's edge, $u_r^{\mathrm{axi}}(W)$, satisfies
\begin{equation}
\frac{u_r^{\mathrm{axi}}(W)}{W}=\frac{\stt(W)-\nu\srr(W)}{Y}=(1-\nu)\beta -\frac{1}{8}\tW^2.
\label{eqn:ur_edge}
\end{equation}

Crucially, \eqref{eqn:sttLaminated} determines the numerical prefactor in the scaling~%$\tWcomp\sim\beta^{1/2}$ of 
(\ref{eq:crit-2}), namely:
%shows that when $\tW$ becomes sufficiently large, in particular when
\begin{equation}
\tWcomp = 2^{3/2}\beta^{1/2} \ , 
%\tW>\tW_\ast=2^{3/2}\beta^{1/2},
\label{eqn:WcHighBend}
\end{equation}
so that for $\tW >\tWcomp$   %then 
the hoop stress becomes compressive at the edge of the sheet, $\sttmin<0$. 
As we discussed in the introduction, this scaling matches the experimentally determined critical sheet size at the onset of delamination for the thinnest elastic sheets (see fig.~\ref{fig:InitialRegimeDiagram}E).

\begin{figure}
\begin{center}
\includegraphics[width=.85\linewidth]{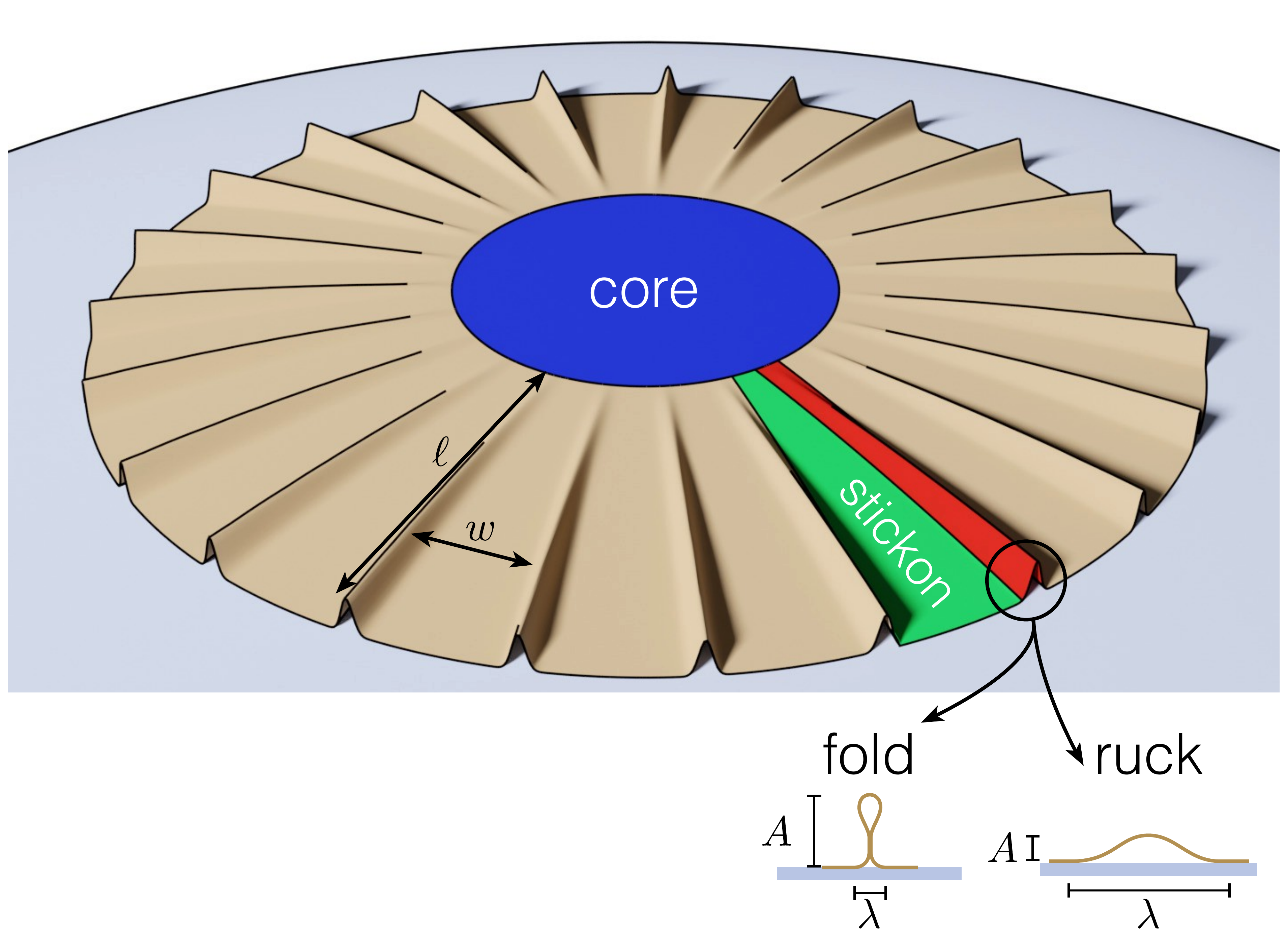}
\end{center}
\caption{Schematic of a thin sheet attached to a sphere highlighting the adhered core region (blue), delamination blisters (red) and adhered ribbons, or `stickons', between the blisters (green). Depending on the relative strength of bending and adhesion in this problem, the delamination blisters may form either folds or rucks (shown below the main figure). The length, $\ell$, of the blisters, as well as the width $w$ of the stickon region are also shown.}
\label{fig:scheme_highlights}
\end{figure}

To understand how and when the appearance of compression affects the onset of delamination, we begin by considering the case in which the sheet has very little resistance to bending. In this case, a recent study of the one-dimensional analogue problem \cite{Davidovitch2021}  suggests that delamination blisters in highly bendable sheets take the form of `folds': the amplitude $A$ is large in comparison with the width of the blister $\lambda$ (as shown in fig.~\ref{fig:scheme_highlights}) and, further, that $\lambda\sim\lbc$ where $\lbc=(B/\Gamma)^{1/2}$ is the \emph{bendocapillary length} and $B\propto Et^3$ is the bending stiffness of the sheet. In the axisymmetric case we similarly  expect  folds to be energetically favorable when $\tW > \tWcomp$ and the bending stiffness is sufficiently small. We therefore define a third dimensionless group, that depends on the bending modulus 
\begin{equation}
\epsilon=\frac{B}{\Gamma R^2}=\left(\frac{\lbc}{R}\right)^2.
\label{eq:epsilon}
\end{equation} Here $\epsilon$ is a dimensionless bending stiffness and $\epsilon^{-1}$ characterizes the degree of ``bendability'' of the adhesive film \cite{Davidovitch2011}.

Assuming that the blisters in a sheet delaminating from a sphere are radially elongated and are hence locally one-dimensional, adopting a similar fold shape, we now turn to understanding when such folds are expected and how many of them should form. This will also allow us to perform a consistency check of the fold ansatz \emph{a posteriori}, and to make the notion of `bendable' more precise.

\subsection{The formation of folds}

The key feature of a fold is that the bending energy is  localized close to the contact region and in the loop --- the majority of the arc length of the material is simply uncurved (albeit vertical) --- and so the bending energy of a single fold of extent $\ell$ (see fig.~\ref{fig:scheme_highlights}) in the radial direction is $B\ell/\lbc=(B\Gamma)^{1/2}\ell$. As a result the bending energy in $n$ folds (all of radial length $\ell$) is
\begin{equation}
\Ubend^\fold\sim n (B\Gamma)^{1/2}\ell.
\label{eqn:UbendFold}
\end{equation} As should be expected, $\Ubend^\fold$ penalizes the creation of more folds; to minimize this energy, the system should have as few blisters as possible. However, forming a small number of blisters is expensive in terms of strain energy because the  portion of the sheet that remains adhered still conforms to the sphere; if there are fewer blisters, the attached ribbon-like elements between the blisters (or `stickons', see fig.~\ref{fig:scheme_highlights}) must be wider and so be more highly strained. The strain energy per unit area of a stickon of width $w$ adhered to a sphere of radius of curvature $R$  is  $\uribbon\sim Y w^4/R^4$ \cite{Majidi2008,Meng2014}; assuming that stickons are wide compared to the blisters formed, $w\gg \lambda$, (i.e.~most of the sheet remains laminated to gain adhesion energy) we have that $w\sim W_c/n$ close to onset. The total strain energy stored in these ribbons, $\Ustrain=\uribbon\times W_c\ell$, is
\begin{equation}
\Ustrain\sim \frac{Y W_c^4}{n^4R^4}\times W_{c} \ell.
\label{eqn:Ustrain}
\end{equation} (Note that there is still a strained, fully-laminated core region, but that this does not play a role in the selection of the number of blisters that are formed.)

As expected, the  energy $\Ustrain$ drives the system to have many blisters, thereby competing with the bending energy $\Ubend^\fold$ to determine the optimal number of folds 
\begin{equation}
\nfold\sim \frac{W_c}{R}(\beta^2\epsilon)^{-1/10}\sim (\beta^3/\epsilon)^{1/10},
\label{eqn:nfold}
\end{equation}
 where we use the assumption that folds form in tandem with the emergence of compression, i.e.
\begin{equation}
\tWc\approx \tWcomp \sim\beta^{1/2},
\label{eqn:ThresholdWfolds}
\end{equation}  in the last equality in \eqref{eqn:nfold}.

To understand when  this regime is expected  experimentally, we note two conditions on the formation of folds. First,  folds are distinguished by being much taller than their width, i.e.~$A\gg\lambda$. To estimate the amplitude of folds, we denote the sheet length to be wasted by these blisters by $\DeltaTot$, so that $A\sim\DeltaTot/\nfold$. The fold ansatz is therefore valid provided that $\DeltaTot/\nfold\gg\lbc$ or
\begin{equation}
\frac{\DeltaTot}{R} \gg(\epsilon^4\beta^3)^{1/10}.
\label{eqn:foldsfirstCond}
\end{equation} 
Second, and simplifying further our analysis by considering the case $\nu=0$, we note that since folds form close to the onset of compression in the sheet, their total arclength, $\DeltaTot$, may be estimated as $\DeltaTot\sim |u_r^{\mathrm{axi}}(W)|$. Recalling that the underlying assumption in the fold regime is that $\tWc=\tWcomp+\delta\tW$ with $\delta\tW\ll\tWcomp\sim\beta^{1/2}$, we can estimate $|u_r^{\mathrm{axi}}(W)/W| \sim \tWcomp\,\delta\tW$ from \eqref{eqn:ur_edge}, so that the inequality~\eqref{eqn:foldsfirstCond}  
becomes $\tWcomp^2\,\delta\tW \gg(\epsilon^4\beta^3)^{1/10}$ or
\begin{equation}
(\epsilon^4\beta^{-7})^{1/10}\ll \delta\tW \ll \beta^{1/2}.
\label{eqn:foldsSecondCond}
\end{equation} (For other values of $\nu$, a detailed calculation, given in the Supplementary Information, shows that \eqref{eqn:foldsSecondCond} holds regardless.) The separation of scales in~(\ref{eqn:foldsSecondCond}) 
is only possible if
\begin{equation}
\epsilon\ll\beta^3.
\label{eqn:FoldCondition}
\end{equation} (Note that to determine the conditions under which folds form it was not necessary to evaluate $\delta\tW$ explicitly, and we hence have not needed to calculate $\DeltaTot$ at all!)

Equation~(\ref{eqn:FoldCondition}) makes precise our earlier statements that folds are expected when the bending stiffness of the sheet is sufficiently small: the appearance of folds requires $\epsilon\ll\beta^3\lll1$. We note also from \eqref{eqn:nfold} that this high bendability regime corresponds automatically to a large number of folds --- just as a large number of wrinkles is associated with the small cost of bending and, consequently, little resistance to compression, this is also the case for folds.

When $\epsilon\gtrsim\beta^3$, the fold ansatz used above is no longer self-consistent since it gives rise to folds for which the typical slope $A/\lambda\lesssim1$. If the slope of the delamination structures formed with $\epsilon/\beta^3\gg1$ were indeed small, they would be ridges \cite{Vella2009,Pocivavsek2018,Guan2022} akin to rucks in rugs \cite{Vella2009prl,Kolinski2009}, rather than the folds we have considered so far.
We therefore turn to consider rucks.

\subsection{The formation of rucks}

Unlike a fold, a small-slope ruck of height $A$ and width $\lambda\gg A$ has a sinusoidal profile $\zeta(x)\approx\tfrac{1}{2}A\bigl[1+\cos(2\pi x/\lambda)\bigr]$. As a result, a ruck `wastes' a length $\Delta=\int_{-\lambda/2}^{\lambda/2}\bigl([1+\zeta'(x)^2]^{1/2}-1\bigr)~\dd x\sim A^2/\lambda$. However, for a  ruck, $A$ and $\lambda$ cannot be chosen independently: a local force balance \cite{Obreimoff1930} (or variational arguments \cite{Majidi2007,Wagner2013}) reveal that these two lengths are constrained by the requirement that the radius of curvature of the blister at the delamination point matches the bendo-capillary length \cite{Obreimoff1930,Majidi2007}, $\lbc$.  This condition gives that $\lambda^2/A\sim\lbc$, which, combined with the wasted length constraint, gives
\begin{equation}
\lambda\sim \Delta^{1/3}\lbc^{2/3},\quad A\sim \Delta^{2/3}\lbc^{1/3}.
\label{eqn:RuckDimensions}
\end{equation} 

These scalings have been derived previously in the context of the `sticky elastica' problem \cite{Wagner2013} in which a blister with a given wasted length $\Delta$ is formed and its dimensions measured. Unlike the sticky elastica problem, however, the length to be wasted in each delamination blister  is not controlled here: 
while \eqref{eqn:ur_edge} implies 
a global constraint for the total amount of length that must be wasted, $\DeltaTot$, there is, as yet, no constraint on the number of blisters that will form, each wasting a length $\Delta=\Delta_{\mathrm{tot}}/n$.

Anticipating that the ruck ansatz will be the appropriate (i.e.~self-consistent) one for moderately bendable sheets (which, based on the fold case, should correspond to $\epsilon/\beta^3\gg1$) we also expect that the threshold sheet size for delamination will be well beyond the critical size at which a hoop compression first develops, i.e.~that $\tW_c\gg\beta^{1/2}$. From \eqref{eqn:ur_edge},  we therefore have that $|u_r(W_c)|\sim R\tW_c^3$. Our working hypothesis is that, upon delamination, this excess length is all wasted by buckling, so that $$\DeltaTot \approx u_r(W_c) \sim R\tWc^3.$$ We therefore repeat the energetic balance argument that allowed us to determine the number of folds in the highly bendable limit: as before,  bending energy seeks to form as few rucks as possible, while the strain in the laminated portions of the sheet drives it to form as many as possible. The important difference with the earlier analysis of folds is that the  bending energy within a ruck is distributed all along its arc length;  we therefore have that the bending energy of all $n$ rucks is
\begin{equation}
\Ubend^\ruck\sim n\times B (A/\lambda^2)^2\times \lambda\ell\sim  \Gamma n^{2/3}\DeltaTot^{1/3}\lbc^{2/3}\ell.
\label{eqn:UbendRuck}
\end{equation} 
The strain energy stored in the laminated portions of the sheet, the stickons, is (in scaling terms) independent  of the form that the delamination blisters take, so that \eqref{eqn:Ustrain} still holds. We can then determine that the optimal number of rucks is
\begin{equation}
\nruck\sim(\beta^3\epsilon)^{-1/14}\frac{W_c}{R}\left(\frac{\DeltaTot}{W_c}\right)^{-1/14}\sim(\beta^3\epsilon)^{-1/14}\tWc^{6/7}.
%\sim(\epsilon\beta^3)^{-1/14}\frac{W^{6/7}}{R^{6/7}}.
\label{eqn:nruck}
\end{equation} 

At this stage, an important difference with the fold case emerges:  we do not \emph{a priori} know the value of $\tW$ at which ruck-like delamination blisters will emerge, and so we cannot use \eqref{eqn:nruck} to determine $\nruck$ in terms of $\epsilon$ and $\beta$. Instead, we determine the threshold $\tW_c$ by the standard buckling criterion, i.e.~by equating the residual hoop 
compression in the delaminated state, $\sigma_{\mathrm{res}}\sim B/\lambda^2$ with $\lambda$ given by \eqref{eqn:RuckDimensions} (see ref.~\cite{Paulsen2016}) and the hoop compression of the axisymmetric (laminated) state, $|\stt^{\mathrm{min}}| \sim Y\tW^2$, which we evaluate from \eqref{eqn:sttLaminated}, recalling that $\tWc \gg \beta^{1/2}$ for rucks. This buckling criterion yields the delamination-into-rucks threshold:       
\begin{equation}
\tWc \sim (\beta^3\epsilon)^{1/12}.
\label{eqn:ThresholdWrucks}
\end{equation} 

We note that the threshold in \eqref{eqn:ThresholdWrucks} reproduces the fold scaling \eqref{eqn:ThresholdWfolds}, i.e.~$\tWc\sim\beta^{1/2}$, as $\epsilon/\beta^3\searrow1$, while for $\epsilon/\beta^3\gg1$ the onset for delamination into rucks occurs at $\tWc\gg\beta^{1/2}$. Similarly, we note that when $\epsilon=O(1)$, \eqref{eqn:ThresholdWrucks} recovers the  prediction of the upper bound \eqref{eqn:WcritClassic} that delamination is favourable when $\tW\gtrsim\beta^{1/4}$.

 As a final consistency check, we note that the aspect ratio of delamination rucks at onset is
\begin{equation}
\frac{A}{\lambda}\sim\left(\frac{\Delta}{\lbc}\right)^{1/3}\sim\left(\frac{\DeltaTot/n}{\lbc}\right)^{1/3}\sim\left(\frac{\beta^3}{\epsilon}\right)^{1/12}.
\label{eqn:AspectRatRucks}
\end{equation}
As expected, for $\epsilon\gg\beta^3$ the aspect ratio of the blisters $A/\lambda\ll1$ and our assumption of small-slope rucks (rather than large-slope folds) is indeed self-consistent.

\section{Experimental measurements of onset}

Combining the two predictions \eqref{eqn:ThresholdWfolds} and \eqref{eqn:ThresholdWrucks}, we have that the critical radius at the onset of delamination scales with $\epsilon$ and $\beta$ as
\begin{equation}
\tWc\sim\begin{cases}
\beta^{1/2} &\epsilon\ll\beta^3,\\
(\epsilon\beta^3)^{1/12} &\beta^3\ll\epsilon\ll 1 \ .
\end{cases}
\label{eqn:Wcrit-0}
\end{equation} 
We note that these two results may alternatively be written:
\begin{equation}
\frac{\tWc}{\beta^{1/2}}\sim\begin{cases}
1&\epsilon/\beta^3\ll 1,\\
(\epsilon/\beta^3)^{1/12}& \epsilon/\beta^3\gg 1.
\end{cases}\label{eqn:WcritCombined}
\end{equation}
Since this form presents different results in terms of the effective bendability of the sheets, $\epsilon/\beta^3$, it is a useful one for reconsidering the experimental data presented in fig.~\ref{fig:InitialRegimeDiagram},  to which we now turn.

\begin{figure}
\centering
 \includegraphics[width=0.9\columnwidth]{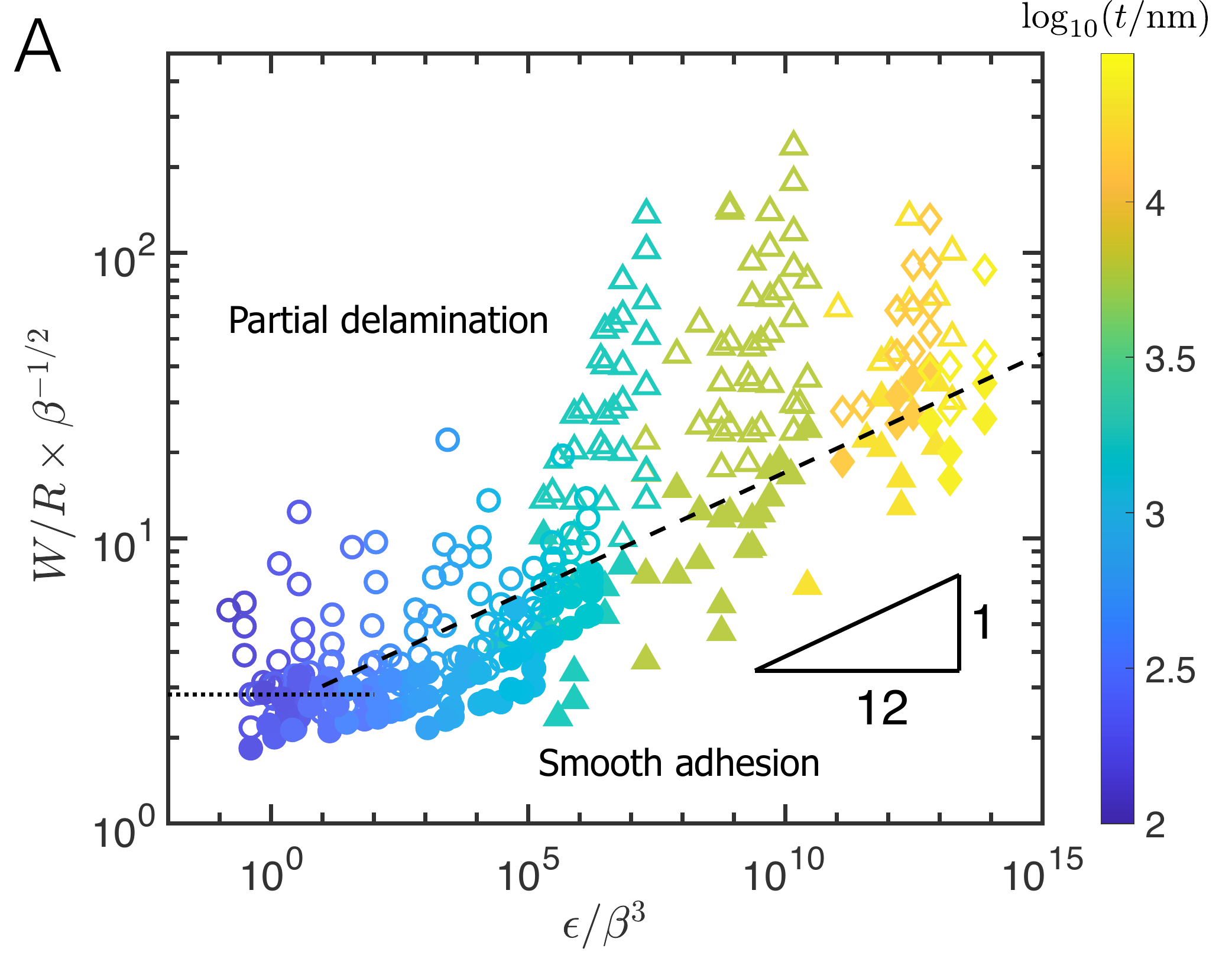}
\includegraphics[width=0.9\columnwidth]{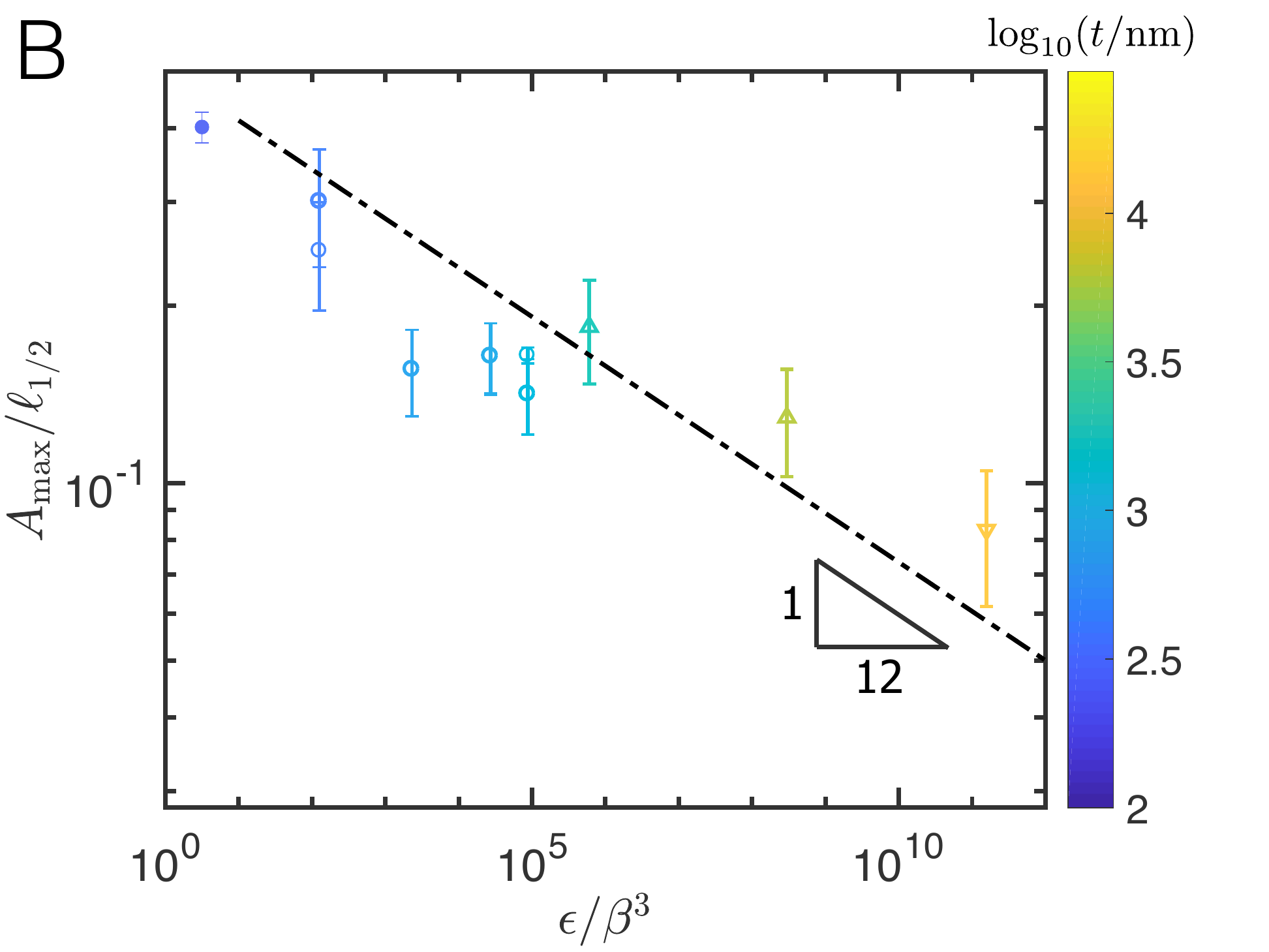}
\caption{Comparison of theory and experiment. A: The regions of parameter space in which smooth adhesion (filled points) and partial delamination (open points) are observed. Here,  data points  are those from fig.~\ref{fig:InitialRegimeDiagram} but  now plotted as suggested by \eqref{eqn:WcritCombined}; this shows a good collapse of the data as well as the two distinct regimes predicted by  theory. (The scaling prediction for rucks, \eqref{eqn:ThresholdWrucks}, is shown by the dash-dotted line; the prediction for folds, \eqref{eqn:ThresholdWfolds}, is shown by the dashed line and corresponds to the dashed line in fig.~\ref{fig:InitialRegimeDiagram}E.)  B: The experimentally measured aspect ratio, showing that experiments are consistent with the scaling for aspect ratio of rucks given in \eqref{eqn:AspectRatRucks} and shown by the dash-dotted line. Here aspect ratios are determined from optical interference (open points) and profilometry (closed point); symbols are used as in fig.~\ref{fig:InitialRegimeDiagram} to show sheet type and thickness.}
\label{fig:RegimesCollapsed}
\end{figure}

As a first comparison between the predictions of the theoretical picture presented above and the experiments already described, we reconsider the data for the state of the system (laminated or delaminated) as a function of sheet radius, presented in fig.~\ref{fig:InitialRegimeDiagram}. Figure \ref{fig:RegimesCollapsed}A shows the data of fig.~\ref{fig:InitialRegimeDiagram} plotted in the way suggested by \eqref{eqn:WcritCombined}; this plot shows that these data collapse well when plotted in this way and, further that the boundary between delaminated and adhered states is consistent with the asymptotic forms predicted in \eqref{eqn:WcritCombined}. (Note that the numerical value of the threshold $\tW_c$ has been fixed to be the value expected in the limit $\epsilon/\beta^3\to0$, $\tW_c=2\sqrt{2}$, by our procedure for determining the adhesion energy $\Gamma$, see Supplementary Information. Nevertheless, the existence of a plateau in this regime is clear, as is the scaling for $\epsilon/\beta^3\gg1$.) We also note that the plateau in $\tWc$ is observed with $\epsilon/\beta^3$ finite but large ($\lesssim100$), rather than strictly $\epsilon/\beta^3\ll1$; this indicates the presence of a large numerical prefactor that cannot be determined by our scaling analysis.

The features of the experimental data as plotted in fig.~\ref{fig:RegimesCollapsed}A are non-trivial tests of the  presented theoretical picture. However, another useful comparison with experiments comes from the aspect ratio of the delamination blisters, measured near the edge of the sheet, in the ruck regime. (This was measured in the case of rucks using an optical interference technique, described in the Supplementary Information, that cannot resolve the large slopes of folds.) These measurements are presented in fig.~\ref{fig:RegimesCollapsed}B and are also consistent with \eqref{eqn:AspectRatRucks}, and hence the theoretical picture as a whole. Note, however, that these experiments were not performed `at' threshold, but at approximately a constant distance beyond it. It is therefore natural to consider the problem of what happens beyond the initial threshold a little further.

\section{Beyond threshold: Spatial structure of folds\label{sec:SpatialVariation}}

The argument so far has focussed on the behaviour at the onset of the delamination instability. However, similar arguments can be used to understand what the desired spatial structure of the fold and ruck pattern might be beyond threshold. Provided that the  radial position $r$ is large compared to the blister width, $r\gg\lambda$, the same arguments used to derive the number of blisters at onset can be repeated with $W_c$ replaced by $r$. In this way we find from \eqref{eqn:nfold} that
\begin{equation}
 \nfold\sim \frac{r}{R}(\beta^2\epsilon)^{-1/10},
\label{eqn:nfoldspatial}
\end{equation} while in the ruck case \eqref{eqn:nruck} gives 
\begin{equation}
n_\ruck(r)\sim (\epsilon\beta^3)^{-1/14}\left(\frac{r}{R}\right)^{6/7}.
\label{eqn:nruckspatial}
\end{equation}

We see that the evolution of  mode number of  instability with radial position, $r$, depends on whether that instability takes place via folds or rucks --- compare the linear scaling with radial position $r$ of \eqref{eqn:nfoldspatial} and the sub-linear scaling with $r$ of \eqref{eqn:nruckspatial}. These two scalings can be written in terms of common variables as:
\begin{equation}
n    \left(\frac{\beta^{3}}{\epsilon}\right)^{1/10}\sim\begin{cases}
    \frac{r}{R}\left(\frac{\beta}{\epsilon^2}\right)^{1/10}&\epsilon\ll\beta^3,\\
\left[\frac{r}{R}\left(\frac{\beta}{\epsilon^2}\right)^{1/10}\right]^{6/7}&\epsilon\gg\beta^3.
    \end{cases}
    \label{eqn:BlisterDistribution}
\end{equation}  

\begin{figure}[h]
\centering
\includegraphics[width=0.95\columnwidth]{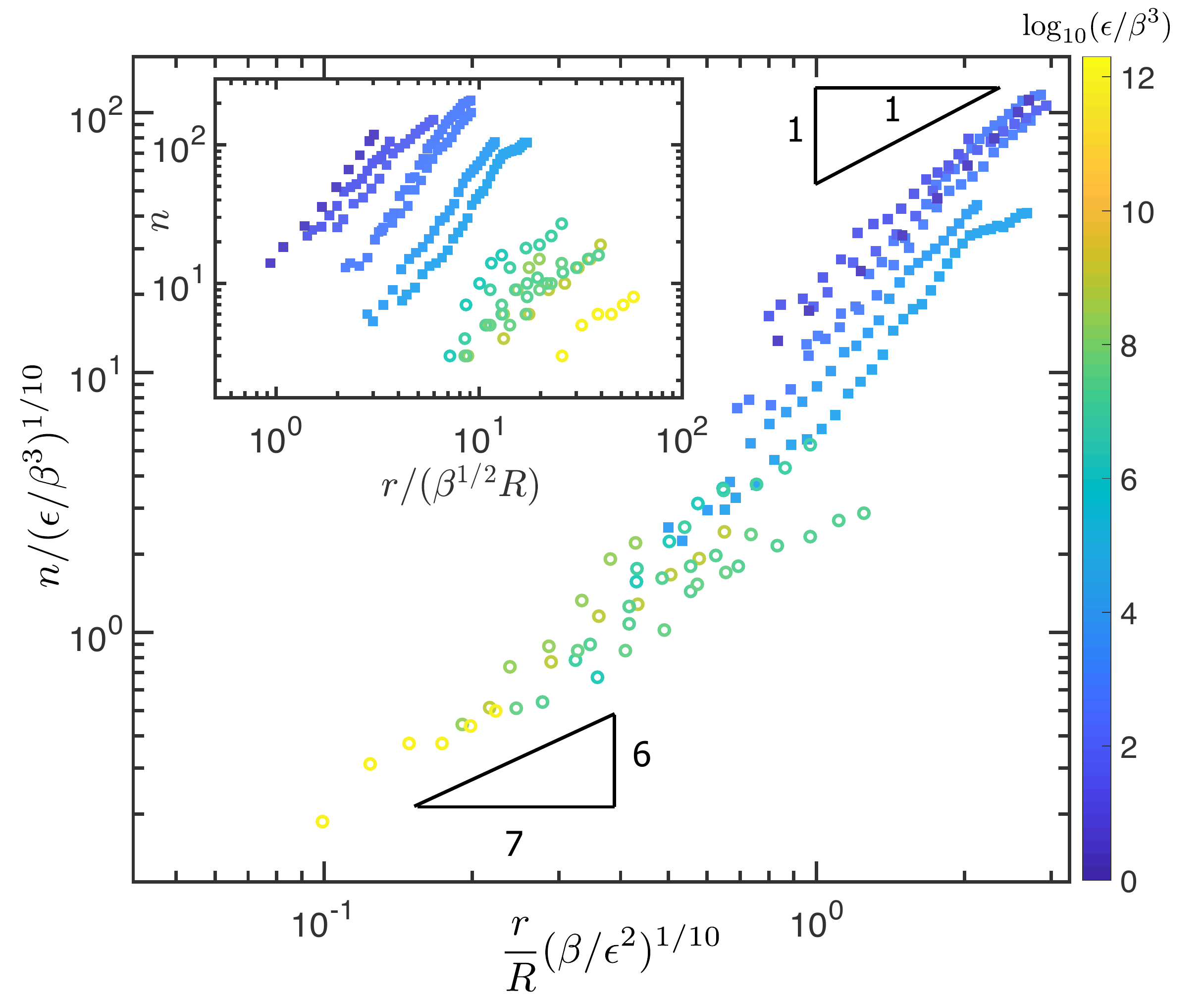}
\caption{The spatial distribution of the fold number, $n$. Here experiments are performed with PS sheets (filled symbols) and PC sheets (open symbols) with the colour bar encoding the value of $\epsilon/\beta^3$. According to the theory, see   \eqref{eqn:BlisterDistribution}, when plotted in this way bluer/darker points should lie closer to the linear scaling expected for folds while yellower/lighter points should lie closer to the $6/7$ exponent expected for  rucks. The experimental results are largely consistent with the theory, and suggest that the exponent of spatial variation  is larger for smaller $\epsilon/\beta^3$. Moreover, the rescaled  data show a reasonable collapse when compared with the raw data (inset).}
\label{fig:NDistribution}
\end{figure}

The predictions of \eqref{eqn:BlisterDistribution} are compared with experimental results in fig.~\ref{fig:NDistribution}: points with larger values of $\epsilon/\beta^3$ (paler/yellower points) are consistent with the sub-linear scaling expected for rucks, while those with smaller values of $\epsilon/\beta^3$ (darker/bluer points) are consistent with the linear scaling expected for folds. The collapse of our experimental data shown in the main portion of fig.~\ref{fig:NDistribution} shows two further noteworthy features: firstly, while the exponents of the two behaviours are ostensibly close ($6/7$ versus unity), the associated prefactors seem to be very different, separating the ruck and fold behaviours; secondly, the steep transition region between the two behaviours suggests that the central regions of highly bendable sheets may form rucks.

\section{Conclusions}

We have shown that the threshold size $\tW_c$ above which a bendable sheet delaminates from an adhesive spherical substrate is not determined merely by the adhesion strength and stretching modulus, as has been supposed previously \cite{Majidi2008,Hure2011,HUre2013,Bense2020}. Instead, 
there is a strong effect of the bending modulus underlying the nontrivial dependence of $\tW_c$ on the two dimensionless groups $\beta$ and $\epsilon$ that is given in \eqref{eqn:Wcrit-0}. 

In particular, we have shown that delamination allows the sheet to expel excess area (allowing conformation) in two different ways. When $\epsilon/\beta^3\ll1$ (or, equivalently, $\Gamma \gg Y \sqrt{t/R}$), the sheet delaminates from the sphere by forming many large-slope folds, while for $\epsilon/\beta^3\gg1$ (i.e.~$\Gamma \ll Y \sqrt{t/R}$), delamination occurs via small-slope rucks.
Despite the differences in the morphology of the delamination pattern, in both cases the onset of instability occurs for sheets that are significantly smaller than the previously presented `upper bound' $\tWdewet \sim\beta^{1/4}$, which was obtained by incorporating the total elastic energy into the standard dewetting criterion for two phases 
\cite{Majidi2008}.

The key difference between the previously identified upper bound on the size of sheets that can conform perfectly to a sphere and our results  arises from the fact that $\tWdewet$ does not take into consideration the 
inhomogeneous, anisotropic distribution of strain in the laminated state, and the consequent presence of compression even for rather small
sheet sizes $\tWcomp <\tW \ll \tWdewet$, where $\tWcomp \sim \beta^{1/2}$, \eqref{eqn:ThresholdWfolds}. Indeed, the emergence of fold and ruck patterns follows directly from the tendency of a thin sheet to maximize adhesion by ``trading'' an energetically-expensive strain with energetically-cheap bending. This principle is analogous to the one underlying wrinkle patterns in sheets confined to curved deformable substrates \cite{Hure2012,King2012,Paulsen2016,Davidovitch2019}; the morphological complexity of delamination patterns in comparison to their wrinkle counterparts follows from the binary (non-analytic) nature of the adhesion energy, which penalizes only the lateral extent of blisters and is insensitive to the deflection amplitude.

We have seen that sheets do not conform perfectly to a spherical substrate once their lateral size increases beyond a limit. In applications that involve a sheet (approximately) conforming to a doubly-curved substrate, it is nevertheless of interest to determine the extent to which some delamination is minor: does the sheet remain largely conformed to the substrate? We therefore define the `conformability' of the sheet to the sphere, $\conf(r)$, as the proportion of the infinitesimal annulus $[r,r+\delta r]$ that is covered by the sheet, i.e.
\begin{equation}
\conf(r)
=1-\frac{n(r)\lambda(r)}{2\pi r}.
\end{equation} Using the results already given we have that the delaminated proportion,  
${\cal D}(r) = 1-\conf(r)$ is
\begin{equation}
{\cal D}(r)\sim\begin{cases}
(\epsilon^2/\beta)^{1/5} & \epsilon\ll\beta^3\ll1,\\
(\epsilon^2/\beta)^{1/7}(r/R)^{4/7} & \beta^3\ll\epsilon\ll1.
\end{cases}
\end{equation} 
The above expressions show three desirable features of the emergence of folds in highly bendable sheets  for uniform conformation:  ${\cal C} \to 1$ in the (singular) infinite bendability limit $\epsilon \to 0$ (i.e.~conformation becomes asymptotically perfect in this limit), ${\cal D}(r)$ is independent of $r$, and, throughout this regime, ${\cal D}\lesssim \beta$. In contrast, the conformation obtained by rucks is nonuniform and less effective in comparison to folds.

We now briefly consider the adhesion of graphene to a spherical substrate, as might be desired in a Surface Force Apparatus, for example. Taking values typical for graphene of $Y\sim 400\mathrm{~N\,m^{-1}}$ \cite{Lee2008} and bending energy $B\sim 1\mathrm{~eV}\sim 10^{-19}\mathrm{~J}$ \cite{Lu2009}, together with a radius of curvature typical of the Surface Force Apparatus, $R\sim 1\mathrm{~cm}$, and an adhesive energy $\Gamma\sim 0.1\mathrm{~N\,m^{-1}}$ we find that $\beta\sim 2\times 10^{-5}$, $\epsilon\sim 10^{-14}$ so that $\epsilon/\beta^3\sim1$, which is sufficient for  graphene to lie within the high bendability regime. In particular, we expect that $\tWc\sim \beta^{1/2}\sim 5\times 10^{-3}$: sheets of graphene adhered to a hemisphere of radius of curvature $1\mathrm{~cm}$ will only adhere smoothly if their radius $W\lesssim 50\mathrm{~\mu m}$. This is a significantly more stringent constraint on the adhesion of graphene to doubly-curved surfaces than would have been the case in the picture presented by the upper bound \cite{Majidi2008}, $\tWdewet\simeq 700~\mathrm{\mu m}$. It may also explain why applications in which graphene spontaneously adheres to curved substrates has been observed to form delamination blisters  \cite{Xu2019} with a morphology similar to the rucks and folds studied here.

The theory and experiments presented in this paper show that the anisotropic and inhomogeneous response of thin elastic sheets to compression manifests itself in both the macroscopic  and the microscopic behaviour of such sheets: both large scale features such as the threshold between delamination and smooth adhesion (not to mention conformability) \emph{and} small scale features like the number of blisters can only be understood through a proper understanding of these different responses. Moreover, this understanding points the way to better control of conformability in scenarios ranging from the humble band-aid to industrial coatings and beyond.

\acknowledgments
The research leading to these results has received funding from  the European Research Council under the European Union's Horizon 2020 Programme / ERC Grant Agreement no.~637334 (DV), the Royal Society   through URF/R1/211730 (FB), the Leverhulme Trust (DV),  the National Science Foundation under grant DMR 1822439 (BD), and a visiting professor fellowship from ENS Lyon (June 2018), where this work was commenced. We are grateful to Daniel Bonn for the use of a profilometer.

\widetext
\clearpage
\begin{center}
\textbf{\large Supplementary information for ``\emph{Delamination from an adhesive sphere: Curvature--induced dewetting versus buckling}"}
\end{center}
%%%%%%%%%% Merge with supplemental materials %%%%%%%%%%
%%%%%%%%%% Prefix a "S" to all equations, figures, tables and reset the counter %%%%%%%%%%
\setcounter{equation}{0}
\setcounter{figure}{0}
\setcounter{table}{0}
\setcounter{section}{0}
\setcounter{page}{1}
\makeatletter
\renewcommand{\theequation}{S\arabic{equation}}
\renewcommand{\thefigure}{S\arabic{figure}}
\renewcommand{\thesection}{S\arabic{section}}

This Supplementary Information file contains further details of the theoretical picture presented in the main paper in \S\ref{sec:Theory}, including the detailed calculation of the length available to waste in the formation of folds. \S\ref{sec:Experiment} provides details of the experimental methods.

\section{Further  theory details \label{sec:Theory}}

\subsection{General calculation of length wasted in folds\label{sec:WastedLength}}

To show that folds can only occur when $\epsilon\ll\beta^3$ (eqn [17] of the main text) it was assumed that the Poisson ratio $\nu=0$. This simplified the calculation of the length that must be wasted by fold formation, $\Delta_{\mathrm{tot}}$,  but the result actually holds for all values of $\nu$. To show this, we consider here how the argument is modified for non-zero Poisson ratios.

The key observation is that for highly bendable sheets, the formation of delamination folds relaxes the compressive stress, \textit{i.e.}~$\sigma_{\theta\theta}\approx0$ \cite{Davidovitch2011}. (More precisely, the maximal level of the compressive stress, $|\min({\sigma_{\theta \theta}})|$, accommodated by the axisymmetric, unbuckled state,  vanishes as $\epsilon \to 0$.) As such, the linear constitutive relation (the tensor equation equivalent of Hooke's law) gives the azimuthal (hoop) strain at the edge of the sheet at threshold as:
\begin{equation}
    \left.\epsilon_{\theta\theta}\right|_{r=W}=\frac{\sigma_{\theta\theta}-\nu\sigma_{rr}}{Y}\approx-\nu\frac{\sigma_{rr}(W)}{Y}=-\nu\beta.
    \label{eqn:epsQQ}
\end{equation}

The azimuthal strain may also be expressed in terms of the radial and normal displacements ($u_r$ and $\zeta$, respectively) as:
\begin{equation}
\epsilon_{\theta\theta}=\frac{u_r}{r}+\frac{1}{2r^2}\left(\frac{\partial\zeta}{\partial\theta}\right)^2.    
\label{eqn:strainDisp}
\end{equation} We therefore combine Eqns [\ref{eqn:epsQQ}] and [\ref{eqn:strainDisp}] with the assumption that in the fold (high bendability) limit, the radial displacement $u_r(W)$ does not change significantly from its value in the axisymmetric setup (\textit{i.e.}~that sufficiently close to threshold, $u_r(W)\approx u_r^{\mathrm{axi}}(W)\cdot\bigl[1+ O(\epsilon^a)\bigr]$ for some $a>0$) to calculate the length to be wasted $\Delta_{\mathrm{tot}}$ by folds, $\Delta_{\mathrm{tot}}=\int_0^{2\pi}\tfrac{1}{2r^2}\left(\tfrac{\partial \zeta}{\partial \theta}\right)^2~W\mathrm{d}\theta $. We find that
\begin{equation}
    -\nu\beta=\left.\epsilon_{\theta\theta}\right|_{r=W}=\frac{u_r(W)}{W}+\frac{\Delta_{\mathrm{tot}}}{W}\approx\frac{u_r^{\mathrm{axi}}(W)}{W}+\frac{\Delta_{\mathrm{tot}}}{W}
\end{equation} and hence
\begin{equation}
    \frac{\Delta_{\mathrm{tot}}}{W}\approx-\frac{u_r^{\mathrm{axi}}(W)}{W}-\nu\beta=\left|\frac{u_r^{\mathrm{axi}}(W)}{W}+\nu\beta\right|.
\end{equation}

Recalling that in the fold regime we anticipate that the critical size for the fold instability is close to that at which compression first occurs, i.e.~$W_c=W_{\mathrm{comp}}+\delta W$ with $\delta W/W_{\mathrm{comp}}\ll1$, and using Eq.~[9] of the main text we then have
\begin{equation}
    \frac{\Delta_{\mathrm{tot}}}{W}\approx\left|\frac{u_r^{\mathrm{axi}}(W)}{W}+\nu\beta\right|=\left|(1-\nu)\beta-\frac{1}{8}\tilde{W}^2+\nu\beta\right|=\left|\beta-\frac{1}{8}\left(2\sqrt{2}\beta^{1/2}+\delta\tilde{W}\right)^2\right|\approx \beta^{1/2}\delta\tilde{W}/\sqrt{2}.
    \label{eqn:GenWastedLength}
\end{equation} \eqref{eqn:GenWastedLength} is identical to the scaling relationship used to derive eqn [17] of the main text under the assumption that $\nu=0$; as a result the main conclusion, eqn [17] of the main text, holds irrespective of Poisson ratio.

\subsection{Leading order correction of $w_c/R$ when $\epsilon/\beta^3\ll1$ \label{sec:threshold}}

The above paragraph highlighted the central role of the compressive stress in the analysis of the highly bendable regime. Namely, it is the hoop component of the axisymmetric stress, $\sigma_{\theta\theta} <0 $,  that is being suppressed upon delamination, rather than the corresponding strain $\epsilon_{\theta\theta}$ (which may retain a finite value $\epsilon_{\theta\theta}\approx -\nu\beta$, even after delamination). Pushing this observation further, we can improve our characterization of the threshold value to delamination folds by equating the maximal compression in the axisymmetric state, $\sigma^{\mathrm{axi}}_{\theta\theta}(r=W)$, with the residual compressive stress in the folded state, $\sigma^{\mathrm{fold}}_{\theta\theta}$. The former is expressed using Eqs~(8) and (10) of the main text together with  $(\tilde{W}_c/\tilde{W}_{\mathrm{comp}})-1 \ll 1$ to give
\begin{equation} \sigma^{\mathrm{axi}}_{\theta\theta}(W) \sim - Y \tilde{W}_{\mathrm{comp}}\cdot(\tilde{W}_c - \tilde{W}_{\mathrm{comp}}). %\sim  - Y \beta^{1/2} (W_c - W_{\mathrm{comp}}) 
\label{eq:SI-qq}
\end{equation} The latter, $\sigma_{\theta\theta}^{\mathrm{fold}}$, is evaluated via the residual (unavoidable) compression in a ``stickon'': $\sigma_{\theta\theta}^{\mathrm{fold}}\sim-Y (w/R)^2 \sim -Y n(W)^{-2} \beta$, where we have used that the width of the delaminated zone is negligible, \textit{i.e.} $\lambda(r) \ll r$, to estimate  the width of each stickon as $w\sim W/n$. Using Eq.~[14] of the main text to estimate $n$, we therefore find that:   
\begin{equation} \sigma^{\mathrm{fold}}_{\theta\theta} 
\sim -Y (\epsilon \beta^2)^{1/5} . 
\label{eq:SI-qq1}
\end{equation}
Combining Eqs.~[\ref{eq:SI-qq}] and (\ref{eq:SI-qq1}) we obtain: 
\begin{equation}
    \tilde{W}_c (\beta, \epsilon) \sim 
      \tilde{W}_{\mathrm{comp}} \cdot \left[1+ c(\epsilon/\beta^3)^{1/5}\right],
\label{eq:threshold}
\end{equation} for some constant $c$. This determines, at a scaling level, the correction to Eq.~[15] of the main text. We shall see below that this prediction is consistent with experimental results performed on the thinnest PS sheets.

\section{Experimental Methods \label{sec:Experiment}}

\subsection{Sheet fabrication and description}

Polystyrene (PS) sheets were fabricated by spin-coating PS (Goodfellow, Cambridge) dissolved in toluene onto glass slides. The thickness of the resulting solid PS sheets was measured using interferometry (Film Metrics F20), and circles of known radius were then cut from the sheet and floated on a bath of tap water. Here, results are presented for sheet thicknesses $120\mathrm{~nm}\leq t\leq 1.55\mathrm{~\mu m}$, with a typical variation of $20\mathrm{~nm}$ within each film.

To reach larger thicknesses (lower bendability), pre-fabricated sheets of other materials were purchased from Goodfellow (Cambridge, UK). In particular, sheets of Polycarbonate (PC) with $t=2\mathrm{~\mu m}, 6\mathrm{~\mu m}$, and Polyimide (PI) with $t=13\mathrm{~\mu m}, 25\mathrm{~\mu m}$ were used.

The geometrical properties of the system (sheet radius $W$, thickness $t$ and sphere radius of curvature $R$) were all known or measured as previously described; similarly the Young's modulus of the sheets is used as reported from the literature, i.e.~$E=3.4\mathrm{~GPa}$ (PS) from \cite{Huang2007}, and $E=2.35\mathrm{~GPa}$ (Polycarbonate)  and $E=2.9~\mathrm{GPa}$ (Polyimide) from Goodfellow data sheets. 

The only unknown in our system is the dry adhesion energy between the sheet and the sphere. This cannot be measured directly using standard blister tests because of the extremely thin sheets used. Instead, we use the behaviour with the very smallest values of $\epsilon/\beta^3$ (where the plateau in $\tilde{W}_c$ is predicted to occur) to fit $\Gamma\approx0.45~\mathrm{~Nm^{-1}}$; more details of this fitting procedure are given in \S1.\ref{sec:AdhStrength}. (It is only possible to measure $\Gamma$ directly for the thinnest sheets, PS, since these  reach the smallest values of $\epsilon/\beta^3$; we therefore  use the same value of $\Gamma$ for all materials.) 

With these parameter values the dimensionless bending stiffness of the sheet, $\epsilon=B/(\Gamma R^2)$, varies in the interval $[2\times10^{-10},0.02]$. 

\subsection{Experimental protocol}

All sheets were floated on the surface of water either as part of the fabrication process (for PS sheets) or manually (for other materials). Commercially available polycarbonate spherical caps were then placed below the floating sheet and the liquid level gradually reduced until the sheet was deposited on the surface of the sphere. The system was then allowed to dry for at least ten minutes before the sphere was inspected to determine whether any delamination had occurred during the drying process. (In some instances, especially close to the transition, sheets delaminated in a portion even though experiments that are nearby in parameter space remained perfectly adhered. Such experiments  are believed to be anomalous, caused by deposition  not  occurring axisymmetrically on the sphere, for example, but are nevertheless indicated by an open point in figures 2 and 4 of the main text for completeness.)

\subsection{Adhesion strength\label{sec:AdhStrength}}

\begin{figure}[h]
    \centering
    \includegraphics[width=0.5\textwidth]{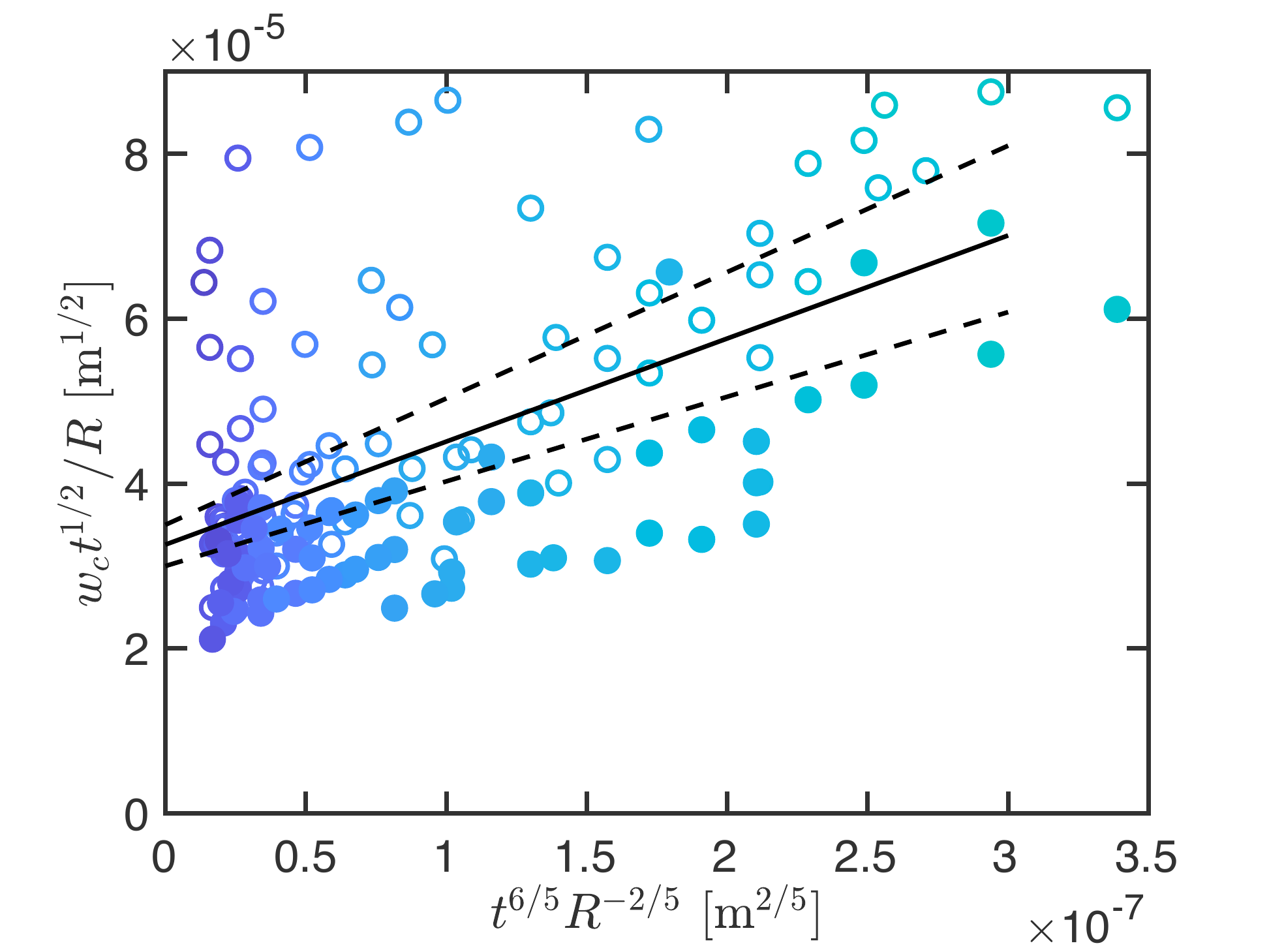}
    \caption{Experimental data obtained with the thinnest PS sheets show that the linear relationship of eqn [\ref{eqn:FitLec}] is observed experimentally. The uncertainty in the value of the intercept is shown by the two dashed lines surrounding the solid line (which corresponds to $\Gamma=0.45\mathrm{~Nm^{-1}}$.}
    \label{fig:FitLec}
\end{figure}

The more detailed version of the theory presented in \S\ref{sec:threshold} gives more  information about how the critical size for delamination is expected to behave with $\epsilon/\beta^3\ll1$; this information can be used in the estimation of the adhesion strength $\Gamma$, as we now discuss. In particular, we expect from \eqref{eq:threshold} that the critical sheet size for delamination with $\epsilon/\beta^3\ll1$ will be:
\begin{equation}
\frac{w_c}{R}=2\sqrt{2}\beta^{1/2}\left[1+c\left(\frac{\epsilon}{\beta^3}\right)^{1/5}\right]
\end{equation} with $c$ undetermined. Now denoting $\ell_{ec}=\Gamma/E$ and recalling that $\beta=\ell_{ec}/t$ we note that this may be rewritten:
\begin{equation}
\frac{w_c}{R}t^{1/2}=2\sqrt{2}\ell_{ec}^{1/2}+c't^{6/5}R^{-2/5}/\ell_{ec}^{3/10}.
\label{eqn:FitLec}
\end{equation} We therefore expect that plotting $w_c\times t^{1/2}/R$ versus $t^{6/5}R^{-2/5}$ will give approximately a straight line with intercept $2\sqrt{2}\ell_{ec}^{1/2}$. Plotting the PS data in this way yields fig.~\ref{fig:FitLec}, which  shows  the linear relationship expected. Fitting this linear relationship, we find that  $\ell_{ec}\approx 0.13\pm0.02\mathrm{~nm}$ for PS (with the   constant $c\approx0.08$). This  corresponds to $\Gamma\approx0.45\pm0.07\mathrm{~Nm^{-1}}$ and so we use the value $\Gamma\approx0.45\mathrm{~Nm^{-1}}$ in all of the plots presented in the main text.

\subsection{Blister shape}

The theory presented throughout the main text is based on the observation that extremely bendable sheets should form folds (rather than rucks) at the onset of the delamination instability. Profiles of delamination blisters around the periphery of the sheet (all located close to the edge of the sheet with $r/W=0.95$) were measured using  a custom made 3D scanning platform that combines a confocal distance sensor (Microepsilon IF2405) with precision linear stages (Physik Instrument), giving (in principle) horizontal and vertical resolutions of $0.5\mathrm{~\mu m}$ and $1\mathrm{~\mu m}$, respectively. We also used a profilometer (Keyence: VK-X1000), reconstructing the profile from the interference pattern observed with light of wavelength $440\mathrm{~nm}$.

While the quantities of interest are the blister width and height ($\lambda$ and $A$ respectively) at an experimental level, it is difficult to measure $\lambda$ directly. Instead we replace $\lambda$ with the blister's full-width at half maximum, $\ell_{1/2}$. Noting that $\Delta/W\sim W^2/R^2$, we then find that in the ruck regime, we should expect:
\begin{equation}
    \frac{A}{\lambda}\sim\left(\frac{\beta}{\epsilon^2}\right)^{1/14}\left(\frac{W}{R}\right)^{5/7}.
    \label{eqn:AspRatFinal}
\end{equation}

Experimental estimates of the typical aspect ratio of the blister, $A/\ell_{1/2}$, measured at the edge of the sheet are plotted in fig.~4B of the main text. These results are consistent with the prediction for the ruck regime \eqref{eqn:AspRatFinal}. (Here,  experiments were performed with a range of values of $\epsilon$, $\beta$ and $\tilde{\Delta}\sim W^2/R^2$. Most experiments were performed with $W/(R\beta^{1/2})= 12\pm1$, corresponding to a horizontal line in the regime diagram of fig.~4A of the main text. However, for the thickest  sheets, this fixed value of $W/(R\beta^{1/2})$ would not lead to delamination;  experiments were therefore performed with sufficiently large $W/(R\beta^{1/2})$ that well-developed delamination occurred.) 

Unfortunately, the interference techniques used did not allow for the measurement of blister profiles in the profiles for experiments with the most bendable sheets (those with the smallest values of $\epsilon/\beta^3$). Nevertheless, the results shown in fig.~4 of the main text are overall consistent with the prediction that more bendable sheets should lead to delamination blisters with larger aspect ratio.

\subsection{Spatial distribution of blisters}

\begin{figure}[h]
    \centering
    \includegraphics[width=0.4\textwidth]{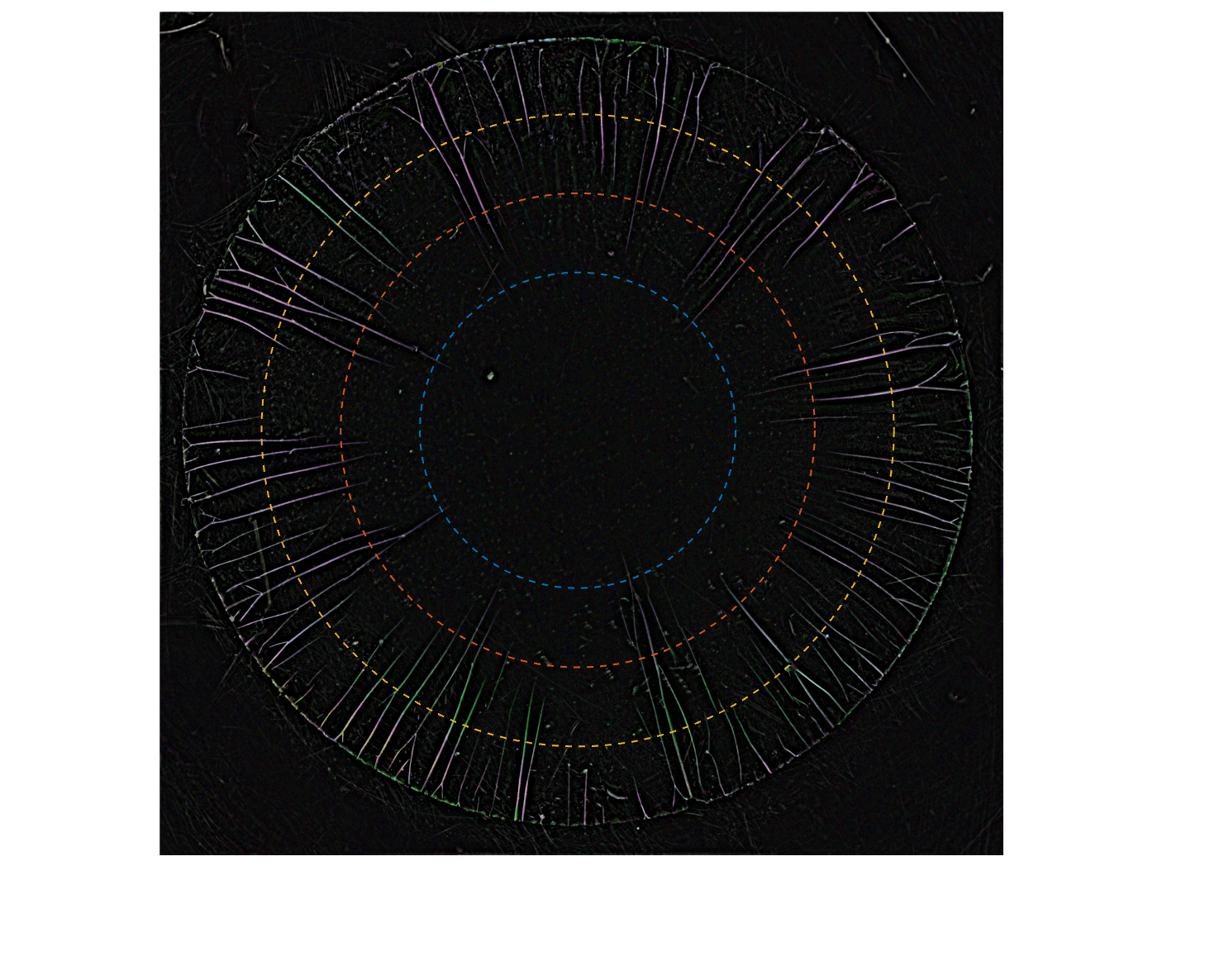}
    \includegraphics[width=0.45\textwidth]{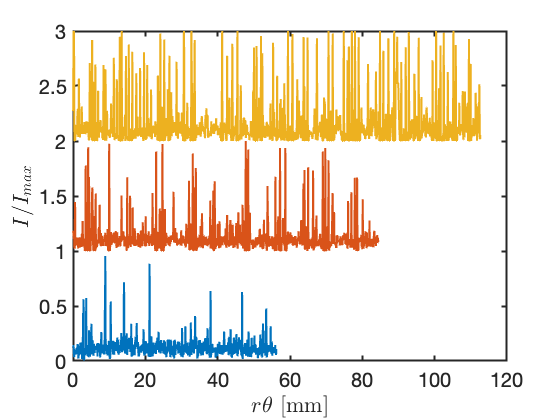}
    \caption{Image detection of blisters at three different radial positions for a PS sheet with $\beta=2.5\times 10^{-4}$, $\epsilon=8.4\times10^{-9}$ and $W/R=0.21$. Left: circles are drawn at various radii (indicated by dashed curves) and the image intensity mapped along this curve. Right: The image intensity along each curve (plotted as a function of arc length) shows a series of peaks; these peaks are counted to give the number of blisters at a given radial position. (Note that the plots in the right hand figure have been offset vertically for clarity.) }
    \label{fig:BlistDistImage}
\end{figure}

For regimes of parameter space in which delamination occurred, a second series of experiments was conducted in which the spatial distribution of the delamination blisters was measured. To do this, images were captured with different focal planes, and a composite image created. (An example of such a composite image is shown in fig.~1D of the main text.) From these images, the  spatial distribution of the number of blisters could be determined by counting peaks in the image intensity in circles of increasing radius --- such intensity peaks correspond to blisters. This process was automated via image processing techniques developed in Matlab: typically, peaks are detected when the peak intensity is more than two standard deviations from the mean of the intensity signal. For some (noisy) images, this threshold was modified to give results that accord with manual counting of peaks. An example of the composite image and the resulting intensity plots is shown in fig.~\ref{fig:BlistDistImage}.

\bibliographystyle{phpf} 
\bibliography{biblio}

\end{document}